\documentclass[a4paper,fleqn,usenatbib,letters]{mnras}

\usepackage{times}
\usepackage[T1]{fontenc}
\usepackage{ae,aecompl}

\usepackage{graphicx}	% Including figure files
\usepackage{amsmath}	% Advanced maths commands
\usepackage{amssymb}	% Extra maths symbols
\usepackage{color}
\usepackage{hyperref}
\newcommand{\xmm}{\textit{XMM-Newton} }
\newcommand{\ns}{\textit{NuSTAR} }

\title[A truncated inner disk in the Rapid Burster]{A strongly truncated inner accretion disk in the Rapid Burster}

\author[Van den Eijnden et al.]{
\noindent J. van den Eijnden$^{1,2}$
\thanks{E-mail: a.j.vandeneijnden@uva.nl},
T. Bagnoli$^{2,3}$,
N. Degenaar$^{1,2}$,
 A. M. Lohfink$^{1}$,
 M. L. Parker$^{1}$,
 \newauthor J. J. M in `t Zand$^{3}$ and A. C. Fabian$^{1}$
\\
$^{1}$Institute of Astronomy, University of Cambridge, Madingley Road, Cambridge CB3 0HA, UK\\
$^{2}$Anton Pannekoek Institute for Astronomy, University of Amsterdam, Science Park 904, 1098 XH Amsterdam, The Netherlands\\
$^{3}$SRON Netherlands Institute for Space Research, Sorbonnelaan 2, 3584 CA Utrecht, the Netherlands
}

\date{Accepted XXX. Received YYY; in original form ZZZ}

\pubyear{2016}

\begin{document}
\label{firstpage}
\pagerange{\pageref{firstpage}--\pageref{lastpage}}
\maketitle

\begin{abstract}
\noindent The neutron star (NS) low-mass X-ray binary (LMXB) the Rapid Burster (RB; MXB 1730-335) uniquely shows both Type-I and Type-II X-ray bursts. The origin of the latter is ill-understood but has been linked to magnetospheric gating of the accretion flow. We present a spectral analysis of simultaneous \textit{Swift}, \ns and \xmm observations of the RB during its 2015 outburst. Although a broad Fe-K line has been observed before, the high quality of our observations allows us to model this line using relativistic reflection models for the first time. We find that the disk is strongly truncated at $41.8^{+6.7}_{-5.3}$ gravitational radii ($\sim 87$ km), which supports magnetospheric Type-II burst models and strongly disfavors models involving instabilities at the innermost stable circular orbit. Assuming that the RB magnetic field indeed truncates the disk, we find $B = (6.2 \pm 1.5) \times 10^8$~G, larger than typically inferred for NS LMXBs. In addition, we find a low inclination ($i = 29\pm2^{\rm o}$). Finally, we comment on the origin of the Comptonized and thermal components in the RB spectrum. 
\end{abstract}


%%%%%%%%%%%%%%%%%%%%%%%%%%%%%%%%%%%%%%%%%%%%%%%%%%

%%%%%%%%%%%%%%%%% BODY OF PAPER %%%%%%%%%%%%%%%%%%

\section{Introduction}

The Rapid Burster (MXB 1730-335, \citealt{lewin76}; hereafter RB) is a peculiar neutron star (NS) low-mass X-ray binary (LMXB) located at a distance of $7.9$ kpc in the globular cluster Liller-1 \citep{valenti10}. NS LMXBs often show X-ray bursts, either due to thermonuclear burning of accreted material on the NS surface (Type-I), or a sudden release of gravitational energy (Type-II). The RB is one of only two NSs showing Type-II X-ray bursts and the only source showing both types. It typically displays only Type-I bursts at high persistent luminosities, and both burst types at lower ones \citep{bagnoli13}. Various models for the poorly understood Type-II bursts have been proposed, including magnetospheric gating of the accretion flow (\citealt{spruit93}, see \citealt{bagnoli15b} for a recent overview of models), in which a strong NS magnetic field truncates the accretion disk outside the innermost stable circular orbit (ISCO; $6$ $R_g$ for a non-spinning NS, where $R_g=GM/c^2$ is the gravitational radius). Measuring the inner disk radius can thus provide a direct test of such magnetospheric models for the RB. 

Constraining the accretion geometry in LMXB is possible by modeling the reflection spectrum \citep{fabian89}: hard X-ray emission reflected off the accretion disk, which prominently contains a gravitationally and dynamically broadended Fe-K line at $\sim 6.5$ keV. Using this approach, \citet{degenaar14} find an inner disk radius of $R_{\rm in} = 85.0 \pm 10.9$ $R_g$ in the Bursting Pulsar (GRO J1744-28, \citealt{kouveliotou96}; hereafter BP), the other source showing Type-II bursts. This truncation is much larger than typically observed in NS LMXBs ($6$--$15$ $R_g$, see e.g. \citealt{cackett10} for a sample study). Hence, the question arises whether a similar truncation is present in the RB. In this Letter, we present an analysis of new, simultaneous observations of the RB using \textit{Swift}, \ns and \textit{XMM-Newton}, aimed at constraining the inner disk radius. 

\section{Observations}

\subsection{\textit{Swift}}

As RB outbursts are relatively predictable, a \textit{Swift} \citep{gehrels04} X-ray Telescope (XRT) monitoring campaign was carried out in Window Timing (WT) mode to detect the start of the outburst. An outburst was detected on 2015 October 3 and triggered, simultaneous \ns and \xmm observations were performed on October 6. A single $\sim500$~s \textit{Swift} observation (obsID 00031360129) coincided with the \ns and \xmm observations. For this observation, we use \textsc{xselect} v2.4d to extract an XRT spectrum from a $70.8$ arcsec radius aperture. We create an arf using \textsc{xrtmkarf}, take the rmf (v15) from the \textsc{caldb}, and rebin the spectrum to ensure a minimum of $20$ counts per bin using \textsc{grppha}. As \textit{Swift}-WT observations of bright, absorbed sources tend to show residuals below $\sim 1$~keV\footnote{See \href{http://www.swift.ac.uk/analysis/xrt/digest_cal.php\#abs}{http://www.swift.ac.uk/analysis/xrt/digest{\_}cal.php{\#}abs}.}, we fit the \textit{Swift} spectrum only in the $1$--$10$ keV range. 

\subsection{\ns}

\ns \citep{harrison13} observed the RB between 2015 October 6 12:11:08 and October 7 15:11:08 (obsID 90101009002), amounting to $\sim46$~ks on-target exposure time for both Focal Plane Modules (FMP) A and B. This full exposure consists of $\sim40$~ks non-burst and $\sim6$~ks burst exposure (see Section \ref{sec:spec} for the determination of burst intervals). We apply the standard routines \textsc{nupipeline} and \textsc{nuproducts} to extract source and background spectra for the non-burst and burst intervals seperately. We extract source spectra from a $120$ arcsec radius circular aperture. As the source dominates its chip, we extract background spectra from a same-sized region on the opposite chip. However, we find no significant differences with background spectra extracted from either the two remaining chips or a smaller background region on the source chip. The source spectrum dominates above the background up to $\sim30$~keV, so we fit the \ns spectra in the range $3$--$30$ keV. 

\subsection{\xmm}

\xmm \citep{jansen01} observed the RB between 2015 October 6 19:23:44 and October 7 06:12:20 (obsID 0770580601) with the EPIC-pn in timing mode and MOS1/2 turned off, resulting in $\sim18$~ks on-target exposure (of which $\sim3$~ks burst exposure). We process the Observation Data Files using \xmm \textsc{SAS} v15. For the RGS detector, we apply \textsc{rgsproc} to extract event lists for non-burst and burst intervals seperately. After assuring no background flaring is present, we combine spectra of the same order from the two RGS detectors using \textsc{rgscombine} and rebin to a minimum of 20 counts per bin. The extracted spectra show discrepancies between the two orders, and between the RGS and XRT spectra, below $1$ and above $2$ keV. Hence, we only consider the range $1$--$2$ keV for the RGS spectra. In this work, we do not search for narrow features in the RGS spectra. 

The EPIC-pn spectrum (extracted using \textsc{epproc} and \textsc{epchain} with quality flag $= 0$, pattern $\leq 4$) shows cross-calibration issues with the \textit{Swift} and \ns spectra: the continuum slope above $\sim5$~keV differs significantly between the spectra. Similar issues in timing mode observations of black-hole binaries in the hard state have been reported recently \citep[see e.g.][]{ingram16fullmod}. Given the consistency between the \textit{Swift} and \ns spectra, we decide to exclude the EPIC-pn spectra. 

\begin{figure}
  \begin{center}
    \includegraphics[width=\columnwidth]{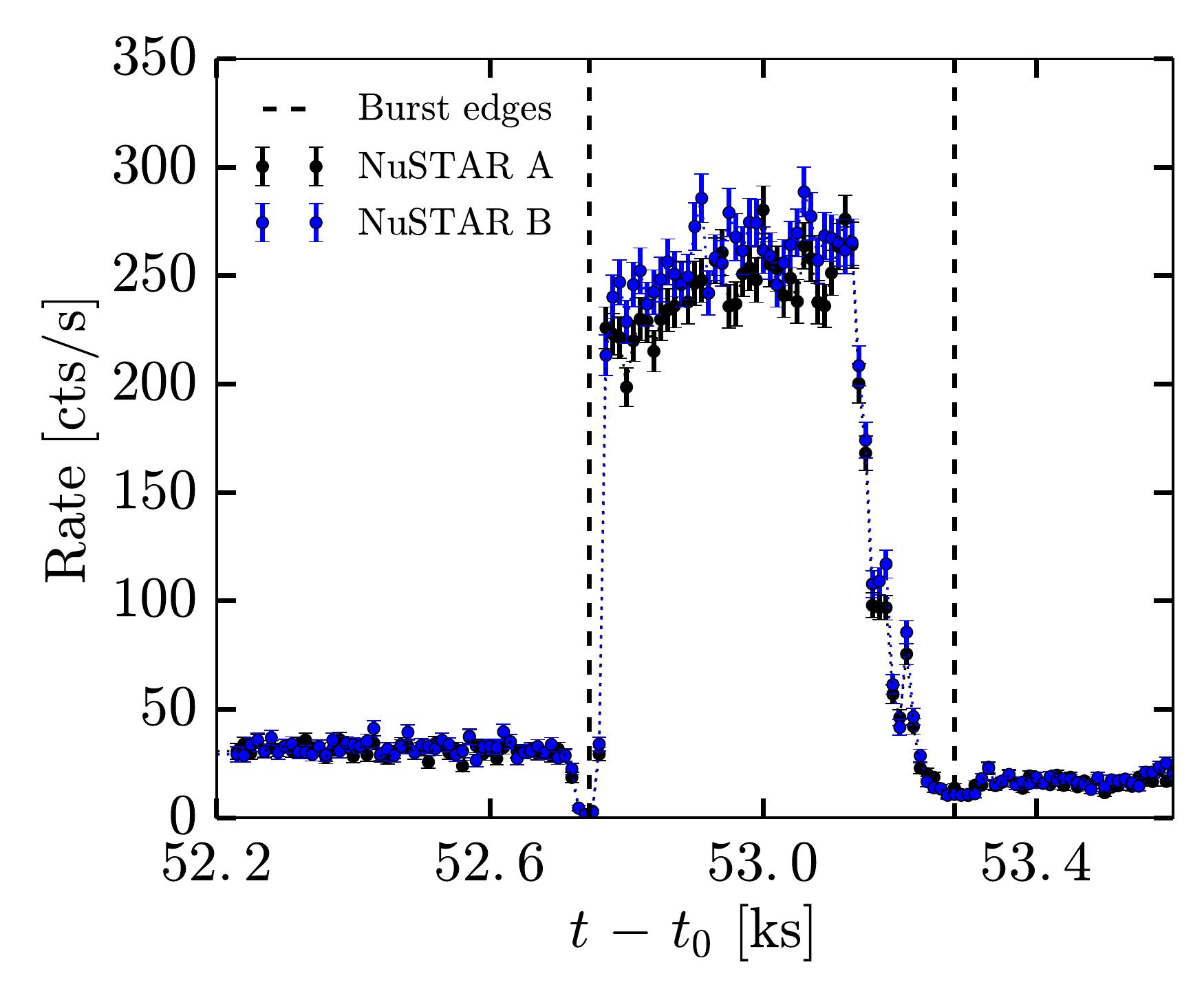}
    \caption{Representative example of a Type-II burst in the \ns observation. The characteric dips, visible before and after the burst, form the border between the non-burst and burst GTIs. $t_0$ is the start time of the observation.}
    \label{fig:lcs}
  \end{center}
\end{figure}

\section{Spectral fitting}
\label{sec:spec}

The \ns and \xmm light curves contain no Type-I bursts and in total 56 Type-II bursts, an example of which is shown in Fig. \ref{fig:lcs}. Measuring the time of the minima in the characteristic dips before and after each burst, we manually define the non-burst and burst intervals used in the extraction of the spectra. Per instrument, we extract a single burst spectrum combining all Type-II bursts. This burst spectrum contains $\sim47\%$ (FMPA,FMPB) and $\sim39\%$ (RGS) of the total counts in the observation. The \textit{Swift} observation coincides fully with a non-burst interval. We initially focus on the non-burst spectrum, and discuss on the burst spectrum at the end of Section \ref{sec:relref}.

We use \textsc{xspec} v12.9.0 \citep{arnaud96} for the spectral fitting and assume solar abundances from \citet{wilms00} and cross-sections from \citet{verner96}. We model interstellar absorption using \textsc{tbabs} and include a free constant between all spectra fixed to $1$ for the FMPA spectrum). All quoted uncertainties are at $1\sigma$.
%All spectra shown in this paper have been rebinned {\color{red} \textbf{in \textsc{xspec} using \textsc{setplot rebin}}}, for visual purposes only

\begin{figure}
  \begin{center}
    \includegraphics[width=\columnwidth]{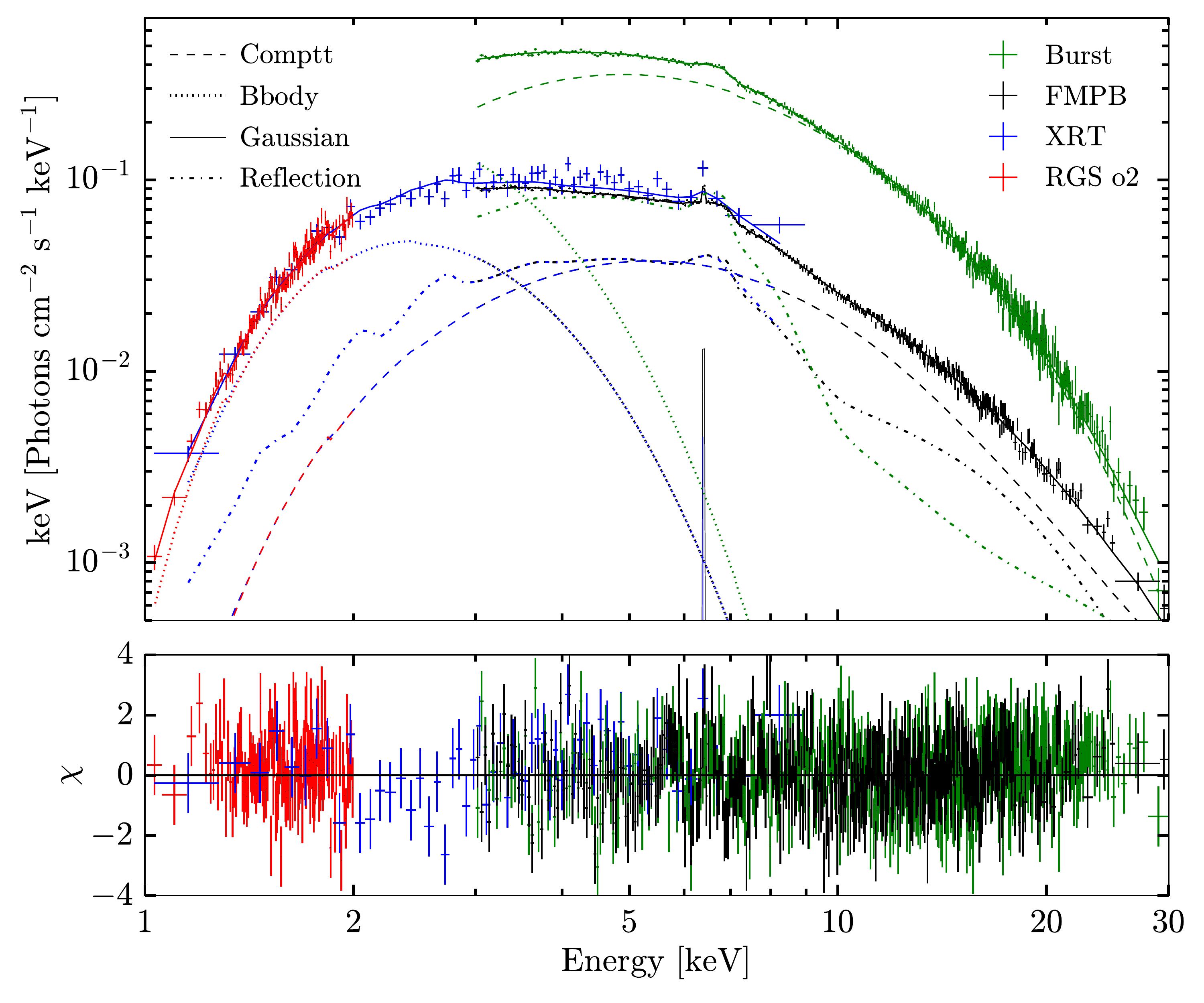}
    \caption{The \ns (FMPB), \xmm RGS (order 2) and \textit{Swift} unfolded spectra with the best fitting \textsc{tbabs} (\textsc{comptt} + \textsc{bbodyrad} + \textsc{relconv*reflionx} + \textsc{gauss}) model. For clarity, we show only a single spectrum per telescope. For comparison, we also show the unfolded FMPA burst spectrum in green. The spectra have been rebinned for visual purposes in \textsc{xspec} using \textsc{setplot rebin}. Note that small deviations appear visible in the \textit{Swift}-spectrum between $2$ and $3$ keV.}
    \label{fig:model}
  \end{center}
\end{figure}

\begin{figure}
  \begin{center}
    \includegraphics[width=\columnwidth]{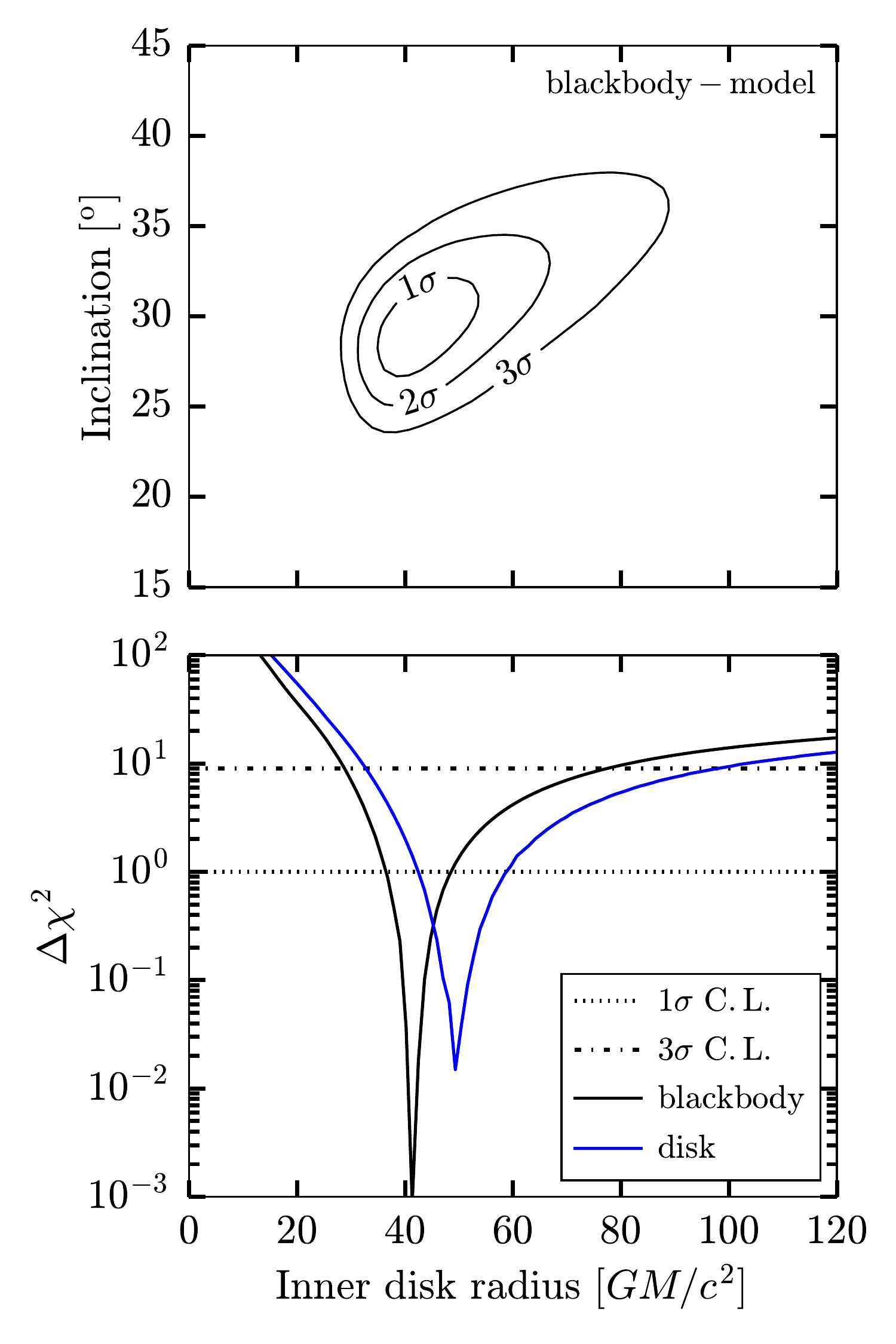}
    \caption{Top: 2D confidence contour for the inclination and inner disk radius or the BB-model. Bottom: confidence plot of the inner disk radius for the blackbody and disk-models.}    
       \label{fig:contour}
  \end{center}
\end{figure}

\subsection{Phenomenological modelling}

\citet{falanga04} fit an \textit{Integral} spectrum ($3$--$100$~keV) of the RB with a model consisting of a powerlaw, a blackbody and a Gaussian Fe-K line. We adopt a similar approach, but replace the powerlaw with the physically-motivated Comptonization model \textsc{comptt} \citep{titarchuk94}. This model yields a reasonable fit with $\chi^2_{\nu}=1.32$ $(3265/2477)$, and the inclusion of the blackbody component is required at high significance ($\Delta \chi^2/\Delta \rm d.o.f = 277/2$). For the Gaussian component, we measure $E_{\rm G} = 6.50\pm 0.02$ keV and $\sigma_{\rm G} = 0.86 \pm 0.03$ keV, consistent with \citet{falanga04}. However, this phenomenological modelling does not provide us with a statistically satisfactory fit, and a Gaussian does not adequately describe the feature around $\sim 6.5$ keV. Hence, we turn to relativistic reflection modelling of the non-burst spectrum. 

\subsection{Relativistic reflection modelling}
\label{sec:relref}
For our reflection fits, we replace the Gaussian line with the model \textsc{reflionx} \citep{ross05}. We apply an adapted version of this model, which was calculated with a \textsc{comptt} illuminating spectrum instead of a powerlaw\footnote{\href{http://www-xray.ast.cam.ac.uk/~mlparker/reflionx_models/reflionx_comptt_hightau.mod}{http://www-xray.ast.cam.ac.uk/$\sim$ mlparker/reflionx\_models/} {\color{blue} reflionx\_comptt\_hightau.mod}}, as \textsc{comptt} is dominant in the phenomenological fit at all energies. To include relativistic effects, we convolve \textsc{reflionx} with \textsc{relconv} \citep{dauser10}. We link the input soft photon temperature $T_0$, electron temperature $kT_e$ and optical depth $\tau$ between \textsc{comptt} and \textsc{reflionx}. We fix the dimensionless spin parameter $a=0.0$, as for NSs this value ranges from $0.0$ to $0.3$ and has little effect on the surrounding metric \citep[see e.g.][]{miller98}. Indeed, setting $a=0.3$ does not yield significant changes in either the model parameters or the quality of the fit. Furthermore, we assume an unbroken emissivity profile with a fixed slope of $q=3$, as the slope is not contrained by the data. This value is consistent with both theoretical expectations \citep{wilkins12} and results from a sample study in NS LMXBs by \citet{cackett10}. The disk ionisation, parametrised as $\xi \equiv 4\pi F/n$, where $F$ is the illuminating flux and $n$ the hydrogen number density, is left variable. In addition, we leave the inclination $i$, inner disk radius $R_{\rm in}$, and iron abundance $A_{\rm Fe}$ free to vary. 

We attempt two possibilities for a soft component: \textsc{diskbb} and \textsc{bbodyrad} (the complete models are hereafter referred to as the disk- and BB-model, respectively). Furthermore, inspection of the residuals around the Fe-K line suggests the presence of an additional narrow emission line around $6.4$ keV. Hence, we also include a narrow Gaussian ($\sigma=10^{-3}$) fixed at this energy of $6.4$ keV. 

The disk-model yields a good fit, with $\chi^2_{\nu}=1.13$ $(2789/2472)$ for a temperature $kT_{\rm disk} = 0.73^{+0.02}_{-0.01}$ keV. The BB-model results in the best fit, with $\chi^2_{\nu}=1.10$ $(2727.8/2472)$ for a temperature $kT_{\rm bb} = 0.55\pm0.01$ keV. All other parameters are listed in Table \ref{tab:pars}. Most interestingly, the reflection component implies a large disk truncation in both models: $R_{\rm in} = 41.8^{+6.7}_{-5.3}$ $R_g$ for the BB-model and $R_{\rm in} = 49.5^{+9.2}_{-7.0}$ $R_g$ for the disk-model, both significantly larger than commonly observed in NS LMXBs \citep[e.g.][]{cackett10}. Both models also yield a consistent, low inclination estimate of $\sim 30^{\rm o}$ and an intermediate disk ionisation of $\xi \sim 470$. The latter is consistent with the typical range observed in both black hole and NS LMXBs ($\log(\xi) \sim 2-3$). The BB-model yields an unabsorped flux between $1$--$30$~keV of $1.47\times10^{-9}$~erg~cm$^{-2}$~s$^{-1}$, corresponding to a luminosity of $1.23\times10^{37}$~erg~s$^{-1}$ (at a distance of $7.9$~kpc) and an Eddington ratio of $\sim 3.2\%$ assuming the emperical Eddington luminosity determined by \citet{kuulkers03}.

Fig. \ref{fig:model} shows the spectra and the best-fitting relativistic reflection $+$ blackbody model. Fig. \ref{fig:contour} shows the confidence contours for the inner disk radius and the inclination. Both parameters are clearly well constrained by the data, and the inner disk radius is inconsistent with the ISCO (i.e. $6$~$R_g$) at $\gtrsim 14.1\sigma$ ($\Delta \chi^2 \gtrsim 200$) for either model. 

The emission line at $6.4$ keV is significant at $\sim 5\sigma$ for both models, given the uncertainty in the normalization. This line, consistent with neutral iron, is also seen in the BP \citep{degenaar14}, although it is more generally observed in high-mass X-ray binaries \citep{torrejon10}. Letting its energy vary does not provide a significant improvement of the fit (f-test probability $p>0.01$ for both models). The residuals in Fig. \ref{fig:model} also suggest the presence of an absorption line around $6.9$ keV. However, the addition of a Gaussian absorption line is not statistically significant.

\begin{table}
 \begin{center}
 \caption{\small{Model parameters for the \textsc{tbabs(comptt} \textsc{+gauss+}soft~component\textsc{+relconv*reflionx)}-models. All quoted uncertainties are at $1\sigma$. We fix $q=3$ and $a=0.0$.}}
  \label{tab:pars}
   \begin{tabular}{llcc}
  \hline \hline
  %Component & Par. [Unit] & Model 1 & Model 2 \\
  Component & Par. [Unit] & BB-model & Disk-model \\
  \hline 
 
  \textsc{tbabs}    & $N_{\rm H}$ [$10^{22}$ $\rm cm^{-2}$] & $3.17 \pm 0.03$               & $3.64 \pm 0.03$ \\
  \textsc{comptt}   & $T_0$ [keV]                           & $1.54 \pm 0.01$               & $1.56 \pm 0.01$ \\
                    & $kT_e$ [keV]                          & $7.19 \pm 0.08$               & $7.12 \pm 0.08$ \\
                    & $\tau$                                & $1.01^{+0.30}_{-0.11}$                  & $1.00^{+0.20}_{-0.11}$ \\
                    & Norm [$10^{-2}$]                      & $1.07\pm0.03$               & $1.10\pm0.03$ \\ 
  \textsc{gauss}    & Norm [$10^{-4}$]                      & $1.66\pm0.35$                              & $1.83\pm0.34$              \\
  \textsc{diskbb}   & $kT_{\rm disk}$ [keV]                 & --                            & $0.73^{+0.02}_{-0.01}$ \\
                    & Norm                                  & --                            & $138\pm15$ \\
  \textsc{bbodyrad} & $kT_{\rm BB}$ [keV]                   & $0.55 \pm 0.01$               & -- \\
                    & Norm                                  & $450\pm40$                    & -- \\
  \textsc{relconv}  & $i$ [$^{\rm o}$]                      & $29\pm2$                      & $32\pm2$ \\
                    & $R_{\rm in}$ [$R_g$]                  & $41.8^{+6.7}_{-5.3}$          & $49.5^{+9.2}_{-7.0}$\\
  \textsc{reflionx} & $\xi$                                 & $470^{+61}_{-14}$                    & $460\pm10$ \\
                    & $A_{\rm Fe}$                          & $0.71^{+0.08}_{-0.06}$                 & $0.77\pm0.06$ \\
                    & Norm                                  & $10.8^{+0.4}_{-0.9}$                  & $10.4^{+0.3}_{-0.5}$ \\
  \hline 
  \end{tabular}
  \end{center}
\end{table}

Fitting the BB-model to the RGS and \ns burst spectra (see Fig. \ref{fig:model}) yields a poorly constrained inclination and $R_{\rm in}$. Thus, we instead fit the burst and non-burst spectra simultaneously, tying the column density and inclination. This results in a good fit ($\chi^2_{\nu} = 1.11$ $(4744.1/4272)$) as now the non-burst data constrains the inclination at $i=22\pm1^{\rm o}$. We measure $R_{\rm in} = 40.1^{+3.3}_{-2.8}$ $R_g$, consistent with the non-burst value. $R_{\rm in}$ is inconsistent with the ISCO at $\gtrsim 8\sigma$. The burst spectrum is significantly softer, yielding a lower electron temperature of $kT_e = 2.42 \pm 0.03$ keV. Finally, the Comptonized component is much stronger relative to the reflection spectrum during burst intervals.
%compared to non-burst

\section{Discussion}

We present a spectral analysis of simultaneous \textit{NuSTAR}, \xmm and \textit{Swift} observations of the RB, aimed at contraining its accretion geometry and the origin of its peculiar Type-II burst behaviour. The non-burst spectrum is well described by a combination of the Comptonization model \textsc{comptt}, relativistic reflection of this Comptonized emission (\textsc{relconv*reflionx}), a soft (disk)blackbody and a narrow emission line at $6.4$ keV. From the reflection spectrum, we measure a large inner disk truncation radius ($\sim 40-50$ $R_g$) and a low inclination ($\sim 30^{\rm o}$). Here, we will discuss the nature of the disk truncation, the implications for Type-II burst models, and the origin of the spectral components. 

\subsection{The nature of the truncated disk}

NS LMXBs show a wide range of inferred inner disk radii, which can roughly be divided into three categories: most sources show small inner disk radii of $6$--$15$~$R_g$ \citep[12 LMXBs, see][]{cackett10, degenaar15, disalvo15, ludlam16, sleator16}. Secondly, five NS LMXBs, mostly (intermittent) X-ray pulsars, show a slightly larger inner radius of roughly $15$--$30$~$R_g$ \citep{iaria16, miller11, papitto13, pintore16, king16}. Finally, two sources show significantly larger truncation radii: \citet{degenaar16a} infer $R_{\rm in} \gtrsim 100$ $R_g$ for IGR J17062-6143, an LMXB persistently accreting a low rates ($L_X \approx 4\times10^{35}$ erg s$^{-1}$). Although various possible explanations exist for a truncated disk at such low $L_X$, truncation by the magnetosphere would imply a large NS magnetic field ($\gtrsim 4\times 10^{8}$ G). Furthermore, \citet{degenaar14} measure $R_{\rm in} = 85.0 \pm 10.9$ $R_g$ in the BP. Our result places the RB in this third category of large disk truncations. As only the RB and the BP show Type-II bursts, the presence of a large disk truncation in both forms an interesting constraint of Type-II burst models.
%Our measurement of its inner disk radius places 
%miller13, 

Models for Type-II bursts can be divided into different general categories \citep[see][for a review]{bagnoli15b}: instabilities in the accretion flow, general-relativistic instabilities close to the ISCO, and interactions between the accretion disk and the NS magnetic field. As only two LMXBs show Type-II bursts, it is difficult to distinguish between these options. However, the first type of model is unable to account for the uniqueness of the RB and the BP: such instabilities should be observed more generally among NS LMXBs. The second category is strongly disfavored by our inner disk radius measurement, as it requires the disk to extend up to the ISCO. 

The most prominent model of accretion instabilities driven by a disk-magnetic field interaction is a so-called \textit{trapped disk} \citep{spruit93,dangelo10}. At the magnetospheric radius $R_m$, the disk is truncated by the magnetic field \citep[see e.g.][]{pringle72}. If the NS spin frequency exceeds the Keplerian rotation frequency of the disk at this radius $R_m$, the magnetic field prevents accretion, confining infalling material in the disk \citep{sunyaev77,spruit93}, or, alternatively, resulting in outflows \citep{illarionov75}. As matter piles up, the magnetospheric radius moves inwards, until the Keplerian frequency exceeds the NS spin, matter swiftly accretes and $R_m$ moves out again. This trapped-disk model is consistent with our measurement of a large truncation radius in the RB, and the similar result for the BP by \citet{degenaar14}. Additionally, in this scenario a change in $R_{\rm in}$ might be expected during bursts. However, the size of such a change in inner radius is unknown; given the relatively large uncertainties on $R_{\rm in}$ ($5$--$10$ $R_g$), we might simply be unable to detect such changes significantly in our observations.

Assuming that the disk is truncated by the magnetic field, we can estimate the magnetic field of the RB using Equation 1 in \citet{cackett09}. We estimate a bolometric flux of $F_{\rm bol} = (2.25\pm0.2) \times 10^{-9}$ erg cm$^{-2}$ s$^{-1}$ by extrapolating the best fit over the $0.1$--$100$ keV range. Using geometrical and efficiency parameters from \citet{cackett09}, we find $B = (6.2 \pm 1.5) \times 10^8 (M/1.4M_{\odot})^2(R/10 \rm km)^{-3}$ G. This estimate is higher than generally observed for NS LMXBs \citep{mukherjee15}. A similarly high magnetic field as in the RB is present in the 11-Hz pulsar IGR J17480-2446 in Terzan 5 \citep{miller11}. This strengthens the proposed link between these two sources \citep{bagnoli13} and indicates that the Rapid Burster, like IGR J17480-2446, could be young, mildly recycled LMXB \citep{patruno12b}. However, a spin measurement for the RB remains necessary to confirm this scenario.

\subsection{Completing the geometrical picture}

\citet{falanga04} fit the continuum in a $3$--$100$ keV \textit{Integral} spectrum with a combination of a powerlaw and a blackbody. The blackbody parameters ($kT_{\rm BB}\sim2.2$ keV, $R_{\rm BB}\sim1.4$ km) suggest that this continuum might originate from a hotspot on the NS surface, as expected if the magnetic field truncates the disk, or a boundery layer. Instead, we fit the continuum above $3$ keV using a \textsc{comptt}-component, which might thus arise from such a hotspot or boundary layer (instead of e.g. a corona). In the trapped-disk model, such an origin could also explain the large increase in the \textsc{comptt}-flux during bursts.%, when presumably the instant mass accretion rate increases strongly.

As we also have high-quality data below $3$ keV, we detect an additional soft blackbody component unseen by \textit{Integral}. Its temperature is consistent with the expectation for the NS surface at the inferred accretion rate \citep{zampieri95}, although the implied radius of $16.7$~km is larger than expected for NSs ($\sim 10$~km). However, \citet{zampieri95} also show that the surface spectrum might slightly deviate from a perfect blackbody. These deviations in the surface spectrum might thus explain why the \textsc{bbodyrad}~model infers a larger radius than expected. If the soft component indeed corresponds to the NS surface, it completes a self-consistent geometrical description of the RB spectrum with a large truncation radius and a hotspot on the NS surface.

\section*{Acknowledgements}

We thank the referee for comments on this Letter. JvdE and ND are supported by a Vidi grant from the Netherlands Organization for Scientific Research (NWO) awarded to ND. ND also acknowledges support via a Marie Curie fellowship (FP-PEOPLE-2013-IEF-627148) from the European Commission. ACF, AL and MP are supported by Advanced Grant Feedback 340442 from the European Research Counsil (ERC). TB acknowledges support from NewCompStar (COST Action MP1304). JvdE and TB acknowledge the hospitality of the Institute of Astronomy in Cambridge, where this research was carried out. We are grateful to Fiona Harrison, Norbert Schartel, Neil Gehrels and the observation planners for making these Director's Discretionary Time observations possible.

%%%%%%%%%%%%%%%%%%%%%%%%%%%%%%%%%%%%%%%%%%%%%%%%%%



%\bibliographystyle{mnras}
%\bibliography{references}
\input{output.bbl}

%%%%%%%%%%%%%%%%%%%%%%%%%%%%%%%%%%%%%%%%%%%%%%%%%%

%%%%%%%%%%%%%%%%% APPENDICES %%%%%%%%%%%%%%%%%%%%%

%\appendix


%%%%%%%%%%%%%%%%%%%%%%%%%%%%%%%%%%%%%%%%%%%%%%%%%%

% Don't change these lines
\bsp	% typesetting comment
\label{lastpage}
\end{document}